\documentclass[aps,prd,nofootinbib,amsmath,amssymb]{revtex4}
\usepackage{eurosym}
\usepackage{amsmath,amssymb}
\usepackage{graphicx}
\usepackage{dcolumn}
\usepackage{bm}
\usepackage{color}
\usepackage{multirow}

\usepackage{float}
\usepackage{array}
\usepackage{dcolumn}
\usepackage{booktabs}
\usepackage[detect-all]{siunitx}
\usepackage{makecell}
\usepackage{xcolor}
\usepackage{yfonts}

\graphicspath{ {/Users/Nastia/Desktop/research_dis/papers_writing/xenes/nov_5} }

\newcolumntype{P}[1]{>{\centering\arraybackslash}p{#1}}
\begin{document}

\title{
Examination of the lattice QCD-motivated strong attractive $\Omega N$  potentials in the $\Omega^- n p$ system}
\author{I. Filikhin$^1$, R. Ya. Kezerashvili$^{2,3,4}$,  and B. Vlahovic$^1$}
\affiliation{\mbox{$^{1}$North Carolina Central University, Durham, NC, USA} \\
$^{2}$New York City College of Technology, The City University of New York,
Brooklyn, NY, USA\\
$^{3}$The Graduate School and University Center, The City University of New
York, New York, NY, USA\\
$^{4}$Long Island University, Brooklyn, NY, USA}

\begin{abstract}
Within the framework of the Faddeev equations in configuration space, we examine the \(\Omega^{-} np\) system, employing strongly attractive lattice HAL QCD and Yukawa-type meson exchange potentials for the \(\Omega N\) interaction.
Our formalism incorporates the attractive Coulomb force between the \(\Omega^{-}\) and proton, treating the system as three non-identical particle pairs (the $ABC$ model). In this study, we assess the impact of the Coulomb interaction on the system and compare our results with recent \(\Omega NN\) ($AAC$ model) calculations, obtained using various approaches. The $ABC$ model yields low-energy characteristics for the \(\Omega NN\) system that differ from previous calculations. 
The Coulomb potential has a marginal perturbative effect on the $AAC$ system, shifting the three-body binding energy by the Coulomb energy of the two-body $BC$ subsystem, but only slightly deviating the spatial configuration from isosceles triangle symmetry. These effects are primarily driven by the strong \(\Omega N\) interaction.
We demonstrate that the large binding energy of the \(\Omega^{-} np\) system arises from the short-range behavior of the \(\Omega N\) potentials.
\end{abstract}

\maketitle

\date{\today }

\section{Introduction}

All quark models, lattice QCD calculations, and other methods predict that
in addition to quark-antiquark mesons and three-quark baryons, there should
be multiquark systems such as dibaryons and tribaryons. A worldwide
theoretical and experimental effort to search for dibaryon states with
strangeness and study strange dibaryon properties has been one of the
long-standing problems in hadron physics. Historically, dibaryons were first
discussed in theoretical studies. In 1977, the possible existence of the $H$
dibaryon ($uuddss$) 
was predicted within the MIT bag
model \cite{Jaffe}. Motivated by the strong attraction between the antikaon
and nucleon the dibaryon with meson-baryon-baryon structure, the $\bar{K%
}NN$ cluster, was predicted in 2002 \cite{Kpp}. Even the system
of three nucleons and antikaons ($\bar{K}NNN$), was the subject of
intensive studies over the last twenty years (See reviews \cite%
{Hyodo2012,Gal2016,Kez10,Kez20}). 
We would also like to mention that since the beginning of the new millennium, studying the composite system from two nucleons and $\Lambda$, $\Xi$, $\Sigma$ or $\phi$ strongly interacting particles has
attracted intense research interest
in many theoretical works  \cite{Miyagawa1995,Filikhin2000,FG2002,BSS,BelyaevLNN,Garcilazo2007,Bel2008,Sofi,Gal2014,GV15,GV2016,GVV16,KamadaLNN2016,FSV17,Gibson2019,Gibson2020,HiyamaXi2020,GV2022,EA24,FKVPRD2024}.  
Unlike the case of
the $NN$ interactions, interactions of these particles with a nucleon are not well determined due to an insufficient number of scattering data. 

Among others di- and tribarions, the
strange $\Omega N$ dibaryon and $\Omega NN$ 
tribarion are the most interesting candidates for study.
The omega baryons, $\Omega $, are a family of hadron particles that are
either neutral or have a +2, +1, or -1 elementary charge. Negatively charged omega,
$\Omega ^{-}(sss)$, are made of three strange quarks \cite%
{B64,DataGroup,2009} and has a rest mass 1672.45 MeV/c$^{2}$ \cite%
{Partlist2012}. It is of particular interest to study the nuclear system with
the strangeness. Due to the strangeness of $\Omega $ baryon, its interactions with nucleons are crucial for understanding the strong force in systems involving heavy strange quarks. 

The $\Omega N$\ dibaryon was predicted to be bound in different quark model
calculations \cite{Goldman1987,Oka1988,Li2000,Pang2004}. For the first time in
Ref. \cite{Goldman1987} was pointed out on the existence of $\Omega N$ 
dibaryon bound state with strangeness $-3$ using the potential quark and MIT
bag models. The possible candidates of $S-$wave dibaryons with various
strange numbers including $\Omega N $ are studied under the chiral $SU(3)$
quark model \cite{Li2000}. The binding energy of the six-quark system with
strangeness $s=-3$ is investigated under the chiral $SU(3)$ constituent
quark model in the framework of Resonating Group Method. The
calculations of the single $\Omega N $ channel with spin $S=2$ are performed.
The effective $\Omega N $ interaction is studied in the refined quark
delocalization color screening model (QDCSM) \cite{Pang2004}. The bound
states are possible because their particular structure has minimal 
contribution from the color-magnetic interaction. Further studies of the
$\Omega N $ dibaryon in the framework of the QDCSM and the chiral quark
model are performed in Refs. \cite{Zhu2015,Huang2015}. 
Although the details in \cite{Goldman1987,Oka1988,Li2000,Pang2004,Zhu2015,Huang2015,SEKIHARA2021} are different, the calculations indicate the existence of the $\Omega N (5/2^{+})$ bound state. 

A lattice QCD
analysis with nearly physical quark masses was performed in Refs. \cite%
{Etminan2014,Morita2016,LagrangianMethod,Iritani2019}. A formalism for
treating the scattering of decuplet baryons in chiral effective field theory
is developed that provides the minimal Lagrangian and potentials in leading-order $SU(3)$
chiral effective field theory for the interactions of octet and decuplet
baryons are provided in Ref. \cite{Meissner2017}. The formalism was applied
for $\Omega N$ and $\Omega \Omega $ scattering, and results were compared
with lattice QCD simulations. Although the details and approaches in studies \cite{Goldman1987,Oka1988,Li2000,Pang2004,Zhu2015,Huang2015,SEKIHARA2021,Etminan2014,Morita2016,LagrangianMethod,Iritani2019} are different, calculations indicate the existence of the $\Omega N$ bound state.

Above, we mentioned the intensive studies of tribaryon clusters
formed by three nucleons and antikaon, and $\Lambda$, $\Xi$, and $\Sigma$ baryons and two nucleons. The bound or resonance state of $\Lambda^{0}$ or $\Sigma ^{-}$,  $\Sigma ^{0}$, $\Sigma ^{+}$ baryons with two nucleons can produce the nuclear system with the strangeness $-1$, while the binding of two nucleons with $\Xi^{-}$ or $\Xi^{0}$ baryons leads to the nuclear system with the strangeness $-2$. What about the formation of tribaryon
nuclear clusters with strangeness $-3$? The concept of a nucleus formed by $\Omega$ baryons and nucleons is an intriguing idea that involves an unusual system where baryons made of strange quarks are bound together with nucleons. To find the lightest $\Omega NN$ system binding energy in Refs. \cite{GV0,GV} used the Faddeev equations in momentum space \cite{GV2016}, where the two-body amplitudes are expanded in terms of Legendre polynomials, and taking into
account that two of the particles are identical. Making use of the $\Omega N$ local potential of Ref. \cite{LagrangianMethod} the authors studied
the $\Omega d$ system the maximal spin channel $(I)J^{P}=(0)5/2^{+}$  \cite{GV0}, while in Ref.  \cite{GV} calculations were performed for $\Omega NN$ system using the HAL QCD Collaboration interaction \cite{Iritani2019}. Employing the same $\Omega N$ potential \cite{Iritani2019} $\Omega NN$ binding energies were calculated in Refs. \cite{Zhang2022,ESE2023} using the method of hyperspherical functions. In Ref. \cite{Zhang2022} the orthogonal basis radial function with one variational parameter was used.

Below, we propose
an investigation of tribaryon cluster $\Omega ^{-}pn$ with strangeness $-3$ in the framework of the method of Faddeev equations in configuration space. The system represents a system of three different pairs. The Coulomb interaction between proton and $\Omega$  baryon yields the consideration of the system as three nonidentical particles ($ABC$ model) instead of the $AAC$ model (the three-body system with two identical particles) which is appropriate without the Coulomb force or using the isotopic spin formalism in which the proton and neutron are the identical particles. The effect of the Coulomb force included in the consideration within the three nonidentical particles formalism is evaluated through numerical analysis by employing HAL QCD interaction \cite{Iritani2019} and the local meson exchange potential \cite{LagrangianMethod}.

We examine the \(\Omega N\) potential using the three-body system \(\Omega^{-} np\). Both \(\Omega N\) potentials~\cite{Iritani2019,LagrangianMethod} are attractive, but yield only a weakly bound \(\Omega N\) pair with a binding energy of 1–2~MeV, which is comparable to the nucleon-nucleon (\(np\)) binding energy of approximately 2.22~MeV. Combining one more neutron to the deuteron results in the formation of the triton, whose ground-state energy is about 3.8 times larger. Previous calculations for the \(\Omega NN\) system have shown that the corresponding ratio is significantly greater—exceeding a factor of ten. Our specific aim is to demonstrate that this pronounced effect in the \(\Omega^{-} NN\) system arises from the short-range behavior of the \(\Omega N\) potential.

This article is organized in the following way. In Sec. \ref{InterPots}, we discuss the $S-$wave HAL QCD $\Omega N$ interaction in  spin-2 channel and the local potential for $\Omega N$($^{5}S_2$) obtained based on a baryon-baryon interaction model with meson exchanges. Faddeev formalism for baryon systems with strangeness $-3$ is presented in Sec. \ref{FaddABC}. Here we present Faddeev equations for the $AAB$ model and $ABC$ model, which includes Coulomb interaction in $\Omega^{-} np$ system. In Sec. \ref{sec:5}, we present and discuss our numerical calculation results, and the summary and concluding remarks follow in Sec. \ref{conclusion}. 

\section{Interaction Potentials}

\label{InterPots}

Investigations of the $\Omega N$ dibaryon states in the strangeness $-3$
channel in Ref. \cite{Etminan2014} authors calculated the $\Omega N$
potential through the equal-time Nambu--Bethe--Salpeter wave function in (2
+1)-flavor lattice QCD with the renormalization group. By solving the Schr%
\"{o}dinger equation with this potential, authors found one bound state with
binding energy 18.9 MeV in state $^{5}S_{2}$. To obtain more accurate and non-perturbative results for the $\Omega$ baryon and nucleon interactions, lattice QCD simulations can be performed. These calculations involve discretizing space and time to simulate the strong interaction in a more controlled way, allowing for predictions about the interaction potentials and scattering amplitudes. Recently, in Ref. \cite{Etminan2014}, the  $\Omega N$ in the $S-$wave and spin-2 channel is studied from
the (2+1)-flavor lattice QCD with nearly physical quark masses ($m_{\pi }$=146 MeV and $m_{K}$=525 MeV) by employing the HAL QCD method. The $\Omega N$($^{5}S_{2}$) potential, obtained under the assumption that its couplings to
the $D-$wave octet-baryon pairs are small, is found to be attractive in all
distances and produces a quasi-bound state 1.54 MeV for $n\Omega ^{-}(uddsss)
$ and 2.46 MeV for $p\Omega ^{-}(uudsss)$. In the later case the binding
energy increase is due to the extra Coulomb attraction. The fitted lattice
QCD potential by Gaussian and Yukawa squared form for obtained observables
such as the scattering phase shifts, root mean square distance, and binding energy, has the form \cite%
{Iritani2019}:
\begin{equation}
V_{L\Omega N}=b_{1}e^{-b_{2}r^{2}}+b_{3}\left( 1-e^{-b_{5}r^{2}}\right)
\left( \frac{e^{-m_{\pi }r}}{r}\right) ^{2}.  \label{HALInteraction}
\end{equation}%
Here the index "L" indicates that it is a lattice QCD potential. Four sets of the fitting parameters $b_{1}$, $b_{2}$, $b_{3}$ and $b_{4}$
are found from the simulation \cite%
{Iritani2019} and the pion mass is $m_{\pi }=146$ MeV. The resultant scattering characteristics obtained with sets of parameters are found to be consistent with each other within statistical errors. The Yukawa squared form at long distance is
motivated by the two-pion exchange between $N$ and $\Omega $.
\begin{figure}[t]
\begin{center}
\includegraphics[width=21pc]{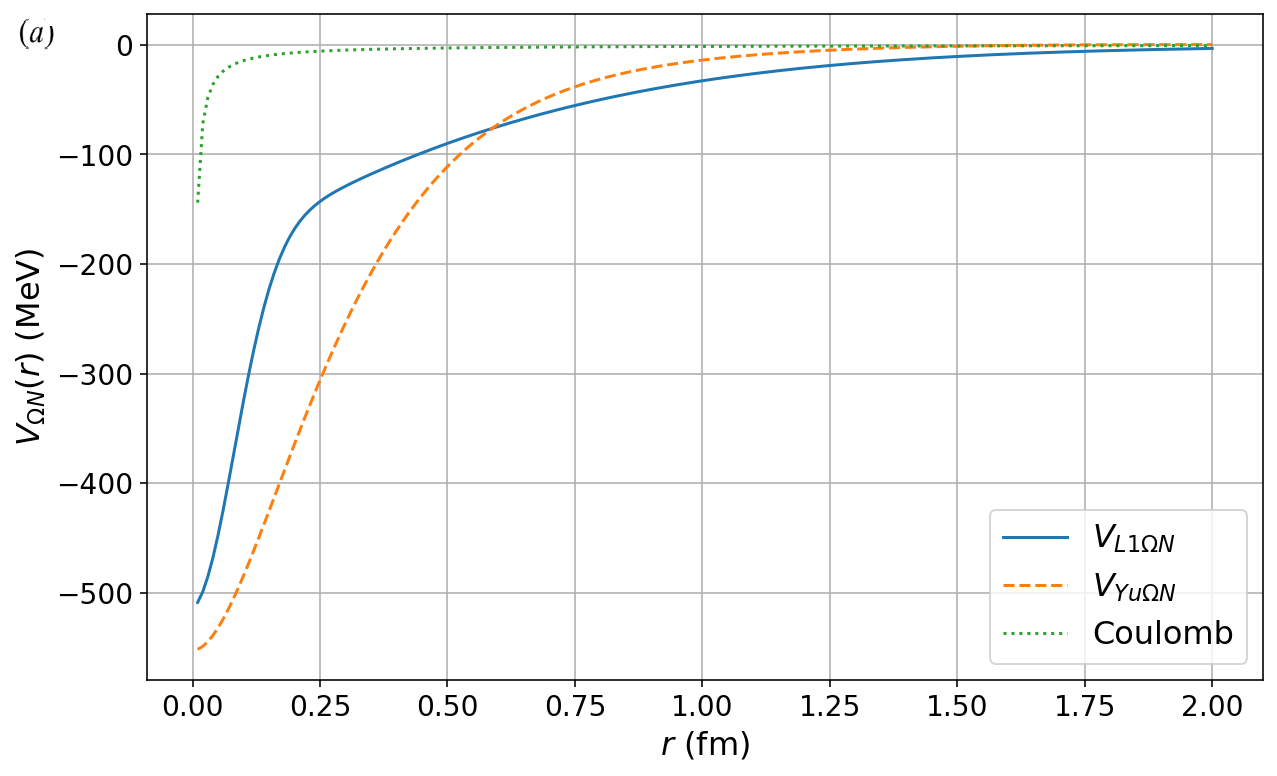}
\includegraphics[width=21pc]{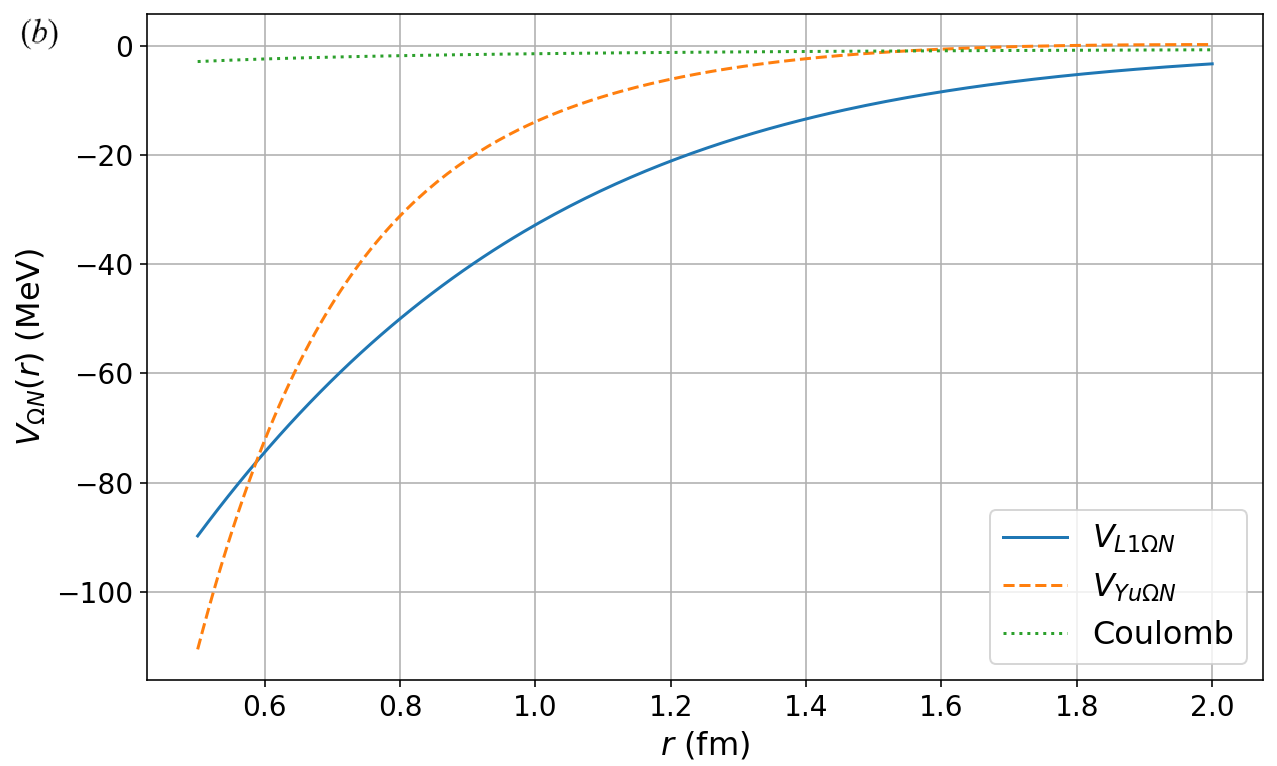}
\end{center}
\caption{The comparison of the Coulomb force and the $\Omega^- N$ potentials near the origin: $(a)$
the region 0.01 fm$< r <2$ fm; $(b)$ the medium-range region 0.5 fm$ < r < 2$ fm.
} \label{fig:41b}
\end{figure}

Based on a baryon-baryon interaction model with meson exchanges in Ref. \cite%
{LagrangianMethod} constructed local potential for the $\Omega N$($^{5}S_{2}$) system
which is useful for calculations. The long-range part of the potential is
related to the exchanges of the $\eta $ meson and of the correlated two
mesons in the scalar-isoscalar channel, while the short-part is represented
by a contact interaction with added inelastic $\Lambda \Xi ,$ $\Sigma \Xi $,
and $\Lambda \Xi $(1530) channels via $K$ meson exchange. The elimination of
these channels induces the energy dependence of the single-channel $\Omega N$
interaction, but this effect is not significant \cite{LagrangianMethod}. These channels effects were assumed to be
small and are neglected in the HAL QCD analyses of the $\Omega N$ 
interaction. The local
potential in coordinate space is expressed as
\begin{equation}
V_{Yu \Omega N}=\frac{1}{2\pi r}\sum_{i=1}^{n}C_{n}\left( \frac{\Lambda ^{2}}{%
\Lambda ^{2}-m_{n}^{2}}\right) ^{2}\left[ e^{-m_{n}r}-\frac{\left( \Lambda
^{2}-m_{n}^{2}\right) r+2\Lambda }{2\Lambda }e^{-\Lambda r}\right],
\label{Hyodo}
\end{equation}%
where $\Lambda =1$ GeV is a cutoff, $m_{n}=n\times 100$ MeV and $C_{n}$ are
the strength of local potential. The values of $C_{i}$\ are given in
\cite{LagrangianMethod}. For a meson exchange $\Omega N$ potential (\ref{Hyodo}) the index "Yu" denotes that it is a Yukawa-type interaction. In Fig. \ref{fig:41b} we present potentials (\ref{HALInteraction}) and (\ref{Hyodo}). The comparison of HAL QCD (\ref{HALInteraction}) and the meson exchange potential (\ref{Hyodo}) shows that at $r < 0.6$ fm potential (\ref{Hyodo}) is wider and stronger than (\ref{HALInteraction}), while at $r  > 0.6$ fm it is less attractive than HAL QCD potential. 
Interestingly enough, the repulsive core is absent in both $\Omega N$ potentials, in
contrast with the nuclear force, because the quark flavors in nucleon
are completely different from those in $\Omega$ baryon and hence the Pauli exclusion principle does not work.

For description of the nuclon-nucleon interaction, we use the MT-I-III \cite{Malfliet1969} and  ATS3 \cite{AT} $NN$ potentials. 

\section{Faddeev formalism for baryon systems with strangeness $-3$ }

\label{FaddABC}

The $\Omega^{-} NN$ is an isospin triplet and there are three
components: $\Omega ^{-}pp(sssuuduud)$, $\Omega ^{-}pn(sssuududd)$, $\Omega
^{-}nn(sssuddudd)$. In addition to the strong interaction, in the $\Omega ^{-}pp$, $\Omega ^{-}pn
$ and $\Omega ^{-}\Omega ^{-}p$ systems, the attractive Coulomb interaction 
between $p$ and $\Omega ^{-}$ will increase the binding energy.

The three-body problem can be solved in the framework of the Schr\"{o}dinger
equation or using the Faddeev approach in the momentum \cite{Fad,Fad1} or
configuration \cite{Noyes1968,Gignoux1974,FM,K86} spaces. The Faddeev equations in the
configuration space have different forms depending on the type of particles
and can be written for: i. three nonidentical particles (\textit{ABC}
model); ii. three particles when two are identical (\textit{AAC} model);
three identical particles (\textit{AAA} model). The identical particles have
the same masses and quantum numbers. The formulation of the Faddeev equations for three particles
can be considered as a starting point of the study of the $\Omega NN$ system. The tribaryon system $\Omega NN$ system can be studied within the \textit{AAC} model with two identical nucleons or by employing the \textit{ABC } model, where two nucleons are distinguishable.

We will use the configuration space formulation for the Faddeev components
of the total wave function. This approach allows us to take into account the
Coulomb force rigorously from the mathematical point of view. In the Faddeev
method in configuration space, which is completely equivalent to finding the wave
function of the three-body system using the Schr\"{o}dinger equation, the
total wave function is decomposed into three components \cite%
{Noyes1968,FM,K86}:
\begin{equation}
\Psi (\mathbf{x}_{1},\mathbf{y}_{1})=\Phi_{1}(\mathbf{x}_{1},\mathbf{y}_{1})+\Phi_{2}(%
\mathbf{x}_{2},\mathbf{y}_{2})+\Phi_{3}(\mathbf{x}_{3},\mathbf{y}_{3}).  \label{P}
\end{equation}%
Each Faddeev component corresponds to a separation of particles into
configurations $(kl)+i$, $i\neq k\neq l=1,2,3$. The Faddeev components are
related to its own set of the Jacobi coordinates ($\mathbf{x}_{i}$, $%
\mathbf{y}_{i}$), $i=1,2,3$. There are three sets of Jacobi coordinates. The
total wave function can be presented by the coordinates of one of the sets as is shown
in Eq. (\ref{P}) for the set  $i=1$. The mass scaled Jacobi coordinates $\mathbf{x}%
_{i}$ and $\mathbf{y}_{i}$ are expressed via the particle coordinates $%
\mathbf{r}_{i}$ and masses $m_{i}$ in the following form:
\begin{equation}
\mathbf{x}_{i}=\sqrt{\frac{2m_{k}m_{l}}{m_{k}+m_{l}}}(\mathbf{r}_{k}-\mathbf{%
r}_{l}),\qquad \mathbf{y}_{i}=\sqrt{\frac{2m_{i}(m_{k}+m_{l})}{%
m_{i}+m_{k}+m_{l}}}(\mathbf{r}_{i}-\frac{m_{k}\mathbf{r}_{k}+m_{l}\mathbf{r}%
_{l})}{m_{k}+m_{l}}).  \label{Jc}
\end{equation}%
In Eq. (\ref{P}), the components depend on the corresponding coordinate set
which are expressed in terms of the chosen set of mass-scaled Jacobi
coordinates. The orthogonal transformation between three different sets of
the Jacobi coordinates has the form:
\begin{equation}
\left(
\begin{array}{c}
\label{tran}\mathbf{x}_{i} \\
\mathbf{y}_{i}%
\end{array}%
\right) =\left(
\begin{array}{cc}
C_{ik} & S_{ik} \\
-S_{ik} & C_{ik}%
\end{array}%
\right) \left(
\begin{array}{c}
\mathbf{x}_{k} \\
\mathbf{y}_{k}%
\end{array}%
\right) ,\ \ C_{ik}^{2}+S_{ik}^{2}=1, \quad k\neq i,
\end{equation}%
where
\begin{equation*}
C_{ik}=-\sqrt{\frac{m_{i}m_{k}}{(M-m_{i})(M-m_{k})}},\quad S_{ik}=(-1)^{k-i}%
\mathrm{sign}(k-i)\sqrt{1-C_{ik}^{2}}.
\end{equation*}%
Here, $M$ is the total mass of the system. 
Let us definite the transformation  $h_{ik}(\mathbf{x},\mathbf{y})$ based on Eq. (\ref{tran}) as 
\begin{equation}
h_{ik}(\mathbf{x},\mathbf{y})=\left(C_{ik} \mathbf{x}+ S_{ik}\mathbf{y}, -S_{ik}\mathbf{x}+ C_{ik}\mathbf{y}
\right). 
\label{Trans}
\end{equation}
The transformation (\ref{Trans}) allows to write the Faddeev equations in compact form. The components $\Phi_i(\mathbf{x}_{i},\mathbf{y}_{i})$ satisfy the Faddeev equations \cite{FM} and can be written in the coordinate
representation as:
\begin{equation}
(H_{0}+V_{i}(C_{ik}\mathbf{x})-E)\Phi_i(\mathbf{x},\mathbf{y})=-V_{i}(C_{ik}\mathbf{%
x})\sum_{l\neq i}\Phi_l(h_{il}(\mathbf{x},\mathbf{y})).
\label{e:1}
\end{equation}
Here, one can choise $k=1$ like to Eq. (\ref{P}), adding the condition  $C_{ii}=1$. $H_{0}=-(\Delta _{\mathbf{x}}+\Delta _{\mathbf{y}})$ is the kinetic
energy operator with $\hbar ^{2}=1$ and $V_{i}(\mathbf{x})$ is the interaction
potential between the pair of particles $(kl)$, where $k,l\neq i$. Equations (\ref{e:1}) presents a system of three coupled second-order differential equations.

Note that we used mass-scaled Jacobi coordinates (\ref{Jc}) for mathematical simplicity. However, physical Jacobi coordinates must be used when calculating physical quantities, such as root mean square distances between particles.  
The distances \( d_\alpha \) (\(\alpha = 1,2,3\)) between the pair $\alpha$ of particles $i$ and $j$ where
$i,j\neq \alpha$ are given as the square root of the expectation value of the square of the non-scaled Jacobi coordinate \( \mathbf{x}_{\alpha} \):  
\begin{equation}  
d_\alpha = \sqrt{\langle \Psi (\mathbf{x}_{\alpha},\mathbf{y}_{\alpha})|\mathbf{x}^2_{\alpha}|\Psi (\mathbf{x}_{\alpha},\mathbf{y}_{\alpha})\rangle }.  
\label{e:dist}  
\end{equation}  
Here, \( (\mathbf{x}_{\alpha},\mathbf{y}_{\alpha}) \) are the non-scaled Jacobi coordinates, and the wave function \( \Psi \) is normalized to 1.  

\subsection{Faddeev equations for \textit{AAC} model}

The system of Eqs. (\ref{e:1}) written for the \textit{ABC} model, can be reduced to a simpler form for a case
of two identical particles, when the particle $B$ in the $ABC$ model is
replaced by the particle $A$. The Faddeev equations in configuration space
for the\textit{\ AAB} model with two identical particles and their application for three-body systems with two identical bosons or fermions are given in our previous
studies \cite{Kez2017,Kez2018PL,KezerasPRD2020}. In the case of two identical fermions,  one must account for the antisymmetrization of the total wave function, and the total wave function of the system is decomposed into the sum
of the Faddeev components $\Phi_1$ and $\Phi_2$ corresponding to the $(AA)B$ and $(AB)B
$ types of rearrangements:
\begin{equation}
\Psi =\Phi_1+\Phi_2-P\Phi_2,
\label{e:00}
\end{equation}%
where $P$ is the permutation operator for two identical fermions.
Therefore, the set of the Faddeev equations (\ref{e:1}) is rewritten as
follows \cite{K86}:
\begin{equation}
\begin{array}{l}
{(H_{0}+V_{AA}-E)\Phi_1=-V_{AA}(\Phi_2-P\Phi_2)}, \\
{(H_{0}+V_{AC}-E)\Phi_2=-V_{AC}(\Phi_1-P\Phi_2)}.%
\end{array}
\label{GrindEQ__1_}
\end{equation}
In Eqs. (\ref{GrindEQ__1_}), ${V_{AA}}$ and ${V_{AC}}$ represent the interaction potentials between identical nucleons, $V_{NN}$, and nonidentical $\Omega$ baryon and nucleon, $V_{\Omega N}$, respectively. 
The spin-isospin variables of the system can be represented by the correspondented basis elements.
After separate of the variables, one can definite 
 the coordinate part, $\Psi^{R}$, of the wave function $\Psi =\xi _{isospin}\otimes \eta _{isospin}\otimes \Psi ^{R}$.

\subsection{Coulomb interaction in $\Omega^- np$ system: $ABC$ model}

The Coulomb force acting between $\Omega^-$ and proton violate the symmetry
of the $AAC$ model. Let us to consider the $ABC$ model, where the
interaction in the pairs $AC$ and $BC$ includes the Coulomb potential.
Below we employing $s$-wave interactions between three particles. A description of the Faddeev equations in configuration space with the Coulomb force is given in \cite{FM,KezerasPRD2020}. The $s$-wave Faddeev equations with the Coulomb interaction that corresponds to the \textit{ABC} model for the $\Omega ^{-}np$ system reads
\begin{equation}
\begin{array}{lll}
(H_{0}+v_{np}+{v}_{C}^{1}-E){\phi_1}&=&-v_{np}({\phi_2}+%
{\phi_3}), \\
(H_{0}+v_{\Omega ^{-}p}+{v}_{C}^{{2}}-E){\phi_2}&=&-v_{\Omega
^{-}p}({\phi_1}+{\phi_3}), \\
(H_{0}+v_{\Omega ^{-}n}+{v}_{C}^{{3}}-E){\phi_3}&=&-v_{\Omega
^{-}n}({\phi_1}+{\phi_2}),%
\end{array}
\label{eq:11a}
\end{equation}%
where
$\phi_i=\Phi^R_i$, $i=1,2,3$  are coordinate parts of the Faddeev components and
\begin{equation}
\begin{array}{l}
v_{C}^{1}=-n/x^{\prime },\quad {v}_{C}^{2}=-n/x,\quad {v}_{C}^{%
3}=-n/x^{\prime \prime },%
\end{array}
\label{eq:11d}
\end{equation}%
 where $n = 1.44$ MeV$\cdot $fm. In Eqs. (\ref%
{eq:11d}) the mass scaled Jacobi coordinate $x^{\prime }=|\mathbf{x_{1}}|$
corresponds to the $\Phi_1$ channel and is expressed by coordinates $x=|%
\mathbf{x_{2}}|$ and $y=|\mathbf{y_{2}}|$ of the channel $\Phi_2$ and $%
x^{\prime \prime }=|\mathbf{x_{3}}|$ is the coordinate of the $\Phi_3$
channel expressed in coordinates $x$ and $y$ of the channel $\Phi_2$
(see Eq. (\ref{e:1}). 
In Eq.~(\ref{eq:11a}), the spin-isospin variables are separated. Formally, this set of equations can be described as \emph{the Faddeev equations for a three-body bosonic system}.

\section{Numerical Results and Discussion}
\label{sec:5}

In our formalism, we are considering the spin and isospin of the particles and assuming that three particles are in $s-$wave by which the spin-isospin state is constructed.
We calculate the eigenenergy (binding energy) of the $\Omega N$ and $\Omega NN$ systems using both \cite{Iritani2019} and \cite{LagrangianMethod} $\Omega N$ local potentials, Malfliet and Tjon (MT) \cite{Malfliet1969} potential for $NN$ interaction and take into account the contribution of the Coulomb potential. We use the $NN$ MT potential  \cite{Malfliet1969} to compare our numerical results with calculations of Refs. \cite{GV0,GV,Zhang2022,ESE2023}, where the same potential was employed. 
For understanding the role of $NN$ interaction on the formation of $\Omega NN$ tribaryon, we also calculate the eigenenergy of this system employing Afnan and Tang (ATS3) \cite{AT} potential. 
To consider the systematic uncertainties from the lattice in results and compare with \cite{GV,Zhang2022,ESE2023} in calculations with the lattice potential (\ref{HALInteraction}) we select two sets of the fitting parameters ($b_{1}$, $b_{2}$, $b_{3}$, $b_{4}$): (--306.5 MeV, 73.9 fm$^{-2}$ --266 Mev fm$^{-2}$, 0.78 fm$^{-2}$) and (--313.0 MeV, 81.7 fm$^{-2}$ --252 Mev fm$^{-2}$, 0.85 fm$^{-2}$). We adopted the following notations for these potentials: $V_{L1\Omega N}$ and $V_{L2\Omega N}$, respectively, while the potential (\ref{Hyodo}) is denoted as $V_{Yu\Omega N}$.

\begin{table}[!ht]
\caption{
The low-energy characteristics of the $\Omega N$ and $np$.  For $\Omega N$ are used the local central HAL QCD  interaction (\ref{HALInteraction}) \cite{Iritani2019} and Yukawa-type meson exchange potential (\ref{Hyodo}) \cite{LagrangianMethod}, while for the spin triplet $np(s=1)$ state, potential \cite{Malfliet1969}. $a_{\Omega N}$, $r_{\Omega N} $, $E_{2}$,  $d$, and  $\Delta_{C}$ are the scattering length, effective range, ground state energy, effective root mean square ($rms$) distance between two particles, and contribution of the Coulomb attraction, respectively.
}
\label{t33}
\begin{tabular}{ccccccc} \hline\noalign{\smallskip}
Potentiall& System & $a_{\Omega N}$ (fm) & $r_{\Omega N} $ (fm)& $E_2$ (MeV)&$d$ (fm) &$\Delta_{C}$, MeV\\
\hline\noalign{\smallskip}
MT \cite{Malfliet1969}& $np(s=1)$         & & & -2.2302& 3.98 &--\\
$V_{L1\Omega N}$ & $\Omega^-n$ & 5.8 & 1.0& -1.2909& 4.11 &--\\
\cite{Iritani2019} & $\Omega^-n$& 5.30&1.26& --&--&-- \\
    & $\Omega^-p$ & 4.8 &0.(9)& -2.1595&3.47& -0.85\\
  \cite{Iritani2019} & $\Omega^-p$&-- &--& -2.18&3.45& $\approx$ -0.9\\
$V_{L2\Omega N}$ & $\Omega^-n$ & 5.7 & 1.2& -1.3757& &--\\
    & $\Omega^-p$ & 4.7 &1.(1)& -2.2645&& -0.89\\
$V_{Yu\Omega N}$ & $\Omega^-n$ &10.2 & 0.75&-0.3385    &-- &--\\
\cite{LagrangianMethod}$^*$ & $\Omega^-n$& 7.4&--& 0.3&3.8&-- \\
      & $\Omega^-p$ &  5.9 & 0.(7)& -1.1794 &--&-0.84\\
   \cite{LagrangianMethod}$^*$ & $\Omega^-p$&5.3&0.75& --&--& $\approx$-0.9\\     
\noalign{\smallskip}\hline
\end{tabular}\\
$^*$the lattice mass values. \ \ \ \ \ \ \ \ \ \ \ \ \ \ \ \ \ \ \ \ \ \ \ \ \ \ \ \ \ \ \ \  \ \ \ \ \ \ \ \ \ \ \ \ \ \ \ \ \ \ \ \ \ 
\end{table}
At the first step we calculate the ground state energies, $E_{2}$, scattering length, $a_{\Omega N}$, the effective range, $r_{\Omega N} $, the effective root mean square distance between two particles, $d$, 
and contribution of the Coulomb attraction, $\Delta_{C}$, for two-particle systems: $NN$ and $\Omega N$. In calculations, the physical masses for $N$ and $\Omega$ listed in Ref. \cite{DataGroup} are used. The corresponding results are presented in Table \ref{t33}. The binding energy for deuteron obtained with the Malfliet-Tjon $^{3}S_{1}$ $NN$ potential \cite{Malfliet1969} is 2.2302 MeV. The contribution of the Coulomb attraction to the binding energy via the lattice QCD potentials \cite{Iritani2019} with different sets of parameters and the meson exchange potential  \cite{LagrangianMethod} are close enough and is approximately about 0.9 MeV. For both $\Omega N$ potentials, we obtain a larger magnitude of the scattering length than the effective range, and the positive $a_{\Omega N}$ indicates the existence of a shallow quasibound state below the threshold. The ground state energies, scattering length, and root mean square distance for $\Omega N$ obtained with $V_{L1\Omega N}$ and $V_{L2\Omega N}$ interactions are close to each other. This demonstrated that the choice of the set of fitting parameters for the HAL QCD potential does not significantly affect these characteristics, and our calculations confirm the results \cite{Iritani2019}. 

There are significant discrepancies between results for $E_2$ obtained with HAL QCD potential and calculations based on the meson exchange $V_{Yu\Omega N}$ potential. To understand these discrepancies let us compare the potentials. Figure \ref{fig:41b} demonstrates the dependences of the HAL QCD, meson exchange, and Coulomb potentials on the distance between the $\Omega$ baryon and the nucleon. Both  
$V_{L1\Omega N}$ and $V_{Yu\Omega N}$ are strong attractive potentials. However, the former represents a more short-range potential, while the latter corresponds to a medium-range potential (see Fig. \ref{fig:41b}($b$)). 

\begin{figure}[b]
\begin{center}
\includegraphics[width=9pc]{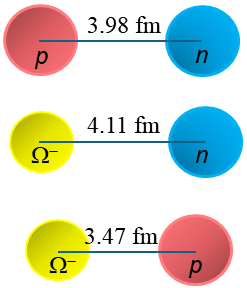}
\end{center}
\caption{The schematics for the $rms$ distances in $pn$, $\Omega n$, and $\Omega p$ pairs, calculated with the $NN$ MT \cite{Malfliet1969} and Hall QCD $V_{L1\Omega N}$ \cite{Iritani2019} potentials. 
The numbers indicate $rms$ distances between the particles. The $\Omega  N$ pairs are differed due to the Coulomb attraction. The sizes of the particles and distances are not in scale.
} \label{fig:2}
\end{figure}

It is interesting to compare distances between particles in the $NN$ and $\Omega N$ systems shown in Fig. \ref{fig:2} calculated with the $NN$ and  $V_{L1\Omega N}$ potentials, respectively. In our formalism, nucleons and $\Omega$ baryon are point-like particles. In Fig. \ref{fig:2}, we use the effective radii of the particles. Modern electron-proton scattering and spectroscopy in muonic hydrogen measurements for the proton $rms$ charge radius are $\sim$ 
0.83–0.87 fm \cite{Protonsize2010,Protonsize2013,Protonsize2017,Protonsize2018,Protonsize2019}, and the neutron has an effective size similar to the proton – roughly $0.8–0.9$ fm \cite{CODATA2018}. The $\Omega^{-}$ composed of three strange quarks is more short‐lived, and its size is less directly accessible experimentally. Lattice QCD and model-dependent calculations \cite{OmegaSize1,OmegaSize2,Gongyo2018} typically yield an $rms$ radius on the order of 0.5–0.7 fm. The most commonly cited value is $\approx$ 0.7 fm. $\Omega^{-}$ has a more compact structure, and its effective radius is somewhat smaller than that of the nucleons. The distances are similar for $np$ and $\Omega^{-} n$ pairs. 
The Coulomb interaction decreases the distance between $\Omega^{-}$ and proton in 
$\Omega^{-} p$ pair.  One can see that the particles are separate for approximately 2 fm. 
In the case of two nucleons, the reason for the separation is a strong repulsive core at short distances for nucleon pair and relatively small binding energy. In the case of $\Omega N$ pair, the same situation does not relate to a repulsive core.
The $\Omega N$ potential has a strong short-range attraction, which does not work to generate large binding energy.
This implies that the differences in \(\Omega N\) potentials are primarily determined by the medium-range interaction. We can evaluate the potential depths corresponding to this inter-particle distance region.

Let us use square quantum well terminology. The minimum depth of the potential well for which the bound state first appears near two-body threshold is given by $U_{0}=\frac {\pi^{2}\hbar^2}{8\mu a^2}$ \cite{Landau}, where $\mu$ is the reduced mass and $a$ is the width of the rectangular quantum well.
Based on Fig. \ref{fig:41b}($a$), one can define the quantum well width $a$ for the nuclear $\Omega N$ potentials as 
$a=1.76$ fm for $V_{L1\Omega N}$ and $a=2.6$ fm for $V_{Yu\Omega N}$. The minimal depth of corresponding restangular
potential is definite as  $U_{01}=-26$ MeV and  $U_{02}=-12$ MeV, respectivily.
For well depth slightly exceeding the minimum value, \textit{i.e.} 
for $U_{0}/U - 1 <<1$ the ground state in $s$-state is given by 
$E=\frac{\pi^2}{16}\frac{(|U_0|-U)^2}{U_0} $ \cite{Landau,Perelomov}. Thus, one can make a fine adjustment for the minimal potential depths to reach the binding energies $B_2$ or the scattering parameters ($a_{\Omega N}$, $r_{\Omega N}$) given in Table. \ref{t33} within the first order of the perturbation theory. 
For example, the value $U_{01}=-8$ MeV results in scattering parameters of $(5.9,\ 1.6)$ fm.
Similarly, modifying $U_{02}$ by $-4$ MeV yields scattering parameters of \((9.9,\ 2.3)\) fm. 
This simple evaluation highlights the effective difference between the potentials $V_{L1\Omega N}$ and $V_{Yu\Omega N}$, which leads to the differences in low-energy characteristics.
Note also that the Coulomb potential remains approximately constant at around 1 MeV in the medium-range region, as demonstrated in Fig. \ref{fig:41b}($b$). This results in a shift of the binding energy by about this value. 

Let us now consider the $AAC$ model for $NN\Omega $ system when the Coulomb interaction is omitted. In Table \ref{tab-2R1} we present the results obtained for the $\Omega d$ state with maximal spin $(I)J^{P} = (0)5/2^{+}$ with different parameter sets for the $\Omega N$ interaction (\ref{HALInteraction}) and the meson exchange potential (\ref{Hyodo}). In calculations are used the physical masses for $N$ and $\Omega$ and masses derived by the HAL QCD Collaboration, 954.7 MeV/c$^{2}$ and 1711.5MeV/c$^{2}$. We calculate the ground state energy $E_{3}$ of $\Omega d$, the three-body energy $E_{3}(V_{np}=0)$, when the interaction
between $n$ and $p$ is omitted, and the energy of the bound $\Omega N$ pair. To find the influence of $NN$ interaction on the formation of the $\Omega N$ we use \cite{Malfliet1969} and \cite{AT} potentials. For comparison, in Table \ref{tab-2R1} are presented the corresponding results from Refs. \cite{GV0,GV}.
\begin{table}[t]
\caption{ The binding energy of the $\Omega d$
 in $AAC$ model when the Coulomb interaction is neglected. Calculations performed with different $\Omega N$ potentials and masses. The mass ratios are 1.78 and 1.79. $E_{3}$ is the ground state energy of the three particle system, $\Omega d$, $E_{2}^{\Omega n}$ is the energy of bound $\Omega N$ pairs, and $E_{3}(V_{NN}=0)$ is the
three-body energy, when the $NN$ interaction between nucleons is omitted, and $\delta$ is the contribution of the mass polarization term. The MT-I-III \cite{Malfliet1969} and  ATS3 \cite{AT} $NN$ potentials are used. The results from Ref. \cite{GV0} are shown for comparison. The energies are given in MeV. 
 }
\label{tab-2R1}
\centering
\begin{tabular}{@{}llccccccc}
\hline
\noalign{\smallskip} Mass, MeV &  Potentials &
$E_{3}$& $E_{3}$\cite{GV0} & $E_{3}(V_{NN}=0)$ & $ E_{2}^{\Omega N}$ &$E_{2}^{\Omega N}$\cite{GV0} & $\delta$\\ \hline
\noalign{\smallskip}
$m_{\Omega}$=1672.45,  &  $V_{L1\Omega N}$, ATS3  & -19.856 & &  -2.997 & -1.2909&-1.29 & 0.4152\\
${m}_{N}$=938.9&  $V_{L1\Omega N}$, MT  & -19.582 &-19.6 & -2.997& -1.2909&-1.29&0.4152\\
        &  $V_{L2\Omega N}$, MT  & -20.010 &-20.0 &  -3.189 & -1.3759&-1.38 & 0.4372\\     
&  $V_{Yu\Omega N}$, MT  & -16.772&-16.34 & -0.974& -0.3385&-0.3&0.2974\\
\hline
\noalign{\smallskip}  
$m_{\Omega}$=1711.50,  & $V_{L1\Omega N}$, MT  & -20.650&-20.6& -3.480 &  -1.5154&-1.52 & 0.4492\\
${m}_{N}$=954.7  & $V_{L2\Omega N}$, MT & -21.099&-21.1& -3.695 &  -1.6112&-1.61 & 0.4726\\
\noalign{\smallskip}
\hline
\end{tabular}%
\end{table}
The analysis of the results in Table \ref{tab-2R1} leads to the following conclusions:

i. The different parameter sets for the $\Omega N$ interaction (\ref{HALInteraction}) and different $NN$ potential change the ground state energy of the $\Omega N$ by about $<$ 0.5 MeV when are used the physical masses for $\Omega$ and $N$. 

ii. Consideration of the masses derived by the HAL QCD Collaboration leads to the ground state energy increase by about 1 MeV. The other energy characteristics listed in Table \ref{tab-2R1} are also increasing.

iii. The omission of the interaction between two nucleons leads to the significant difference for  $E_3(V_{NN}=0)$ calculated with the HAL QCD and meson exchange potentials.

iv. The comparison of the results obtained with the local potentials Eqs. (\ref{HALInteraction}) and (\ref{Hyodo}) shows significant differences for all energy characteristics. Thus, all energy characteristics are very sensitive to the form of the $\Omega N$ interaction. 

v. The comparison of our calculations with Garcilazo and Valcarce \cite{GV0,GV} are in good agreement. 

To analyze the puzzle of difference between $E_3(V_{NN}=0)$ calculated with the HAL QCD and meson exchange potentials, following \cite{H2002} let us use the non
Jacobian form of the Schr\"{o}dinger equation for the $AAC$ model written in the reference frame concerning the nonidentical particle \textit{C}. This equation 
written in a self-explanatory notation reads:  
\begin{equation}
\begin{array}{c}
(-\frac{\hbar ^{2}}{2\mu }\nabla _{r_{A_{1}}}^{2}-\frac{\hbar ^{2}}{2\mu }%
\nabla _{r_{A_{2}}}^{2}-\frac{\hbar ^{2}}{m_{C}}\nabla _{r_{A_{1}}}\nabla
_{r_{A_{2}}}+V_{AA}(r_{A_{1}},r_{A_{2}}) \\ +V_{AC}(r_{A_{1}})  +V_{AC}(r_{A_{2}})-E)\Psi
(r_{A_{1}},r_{A_{2}})=0,  \label{Sh}
\end{array}
\end{equation}%
where $\mu$ is a reduced mass of \textit{A} and \textit{C} particles. In the latter equation the third term is known as the mass polarization term (MPT), $T_{MPT}=-\frac{\hbar^{2}}{m_C}\nabla_{r_{A_1}}\nabla_{r_{A_2}}$. If
 $V_{AA}=0$, then  $%
E\equiv E_{3}(V_{AA}=0)$, which  corresponds to the binding energy of the $AAC$
system when the interaction between two identical particles is omitted.
The mass of each particle $m_A$, $m_C$ is always greater than the reduced mass $\mu $: $ 
m_{C}>m_A>\mu $ and the reduced mass is always less than the mass of
the lightest particle. In the case $m_A>m_{C}$ the contribution of the
MPT can be the same order as the contribution of the
other two differential operators in Eq. (\ref{Sh}). This is due to the comparable
mass factors of these operators, which are approximately $1/m_{C}$. In the
case $m_{C}>m_A$, the contribution of the term $T_{MPT}$ has the factor $1/m_{C}$,
while the mass factors of the other differential operators are the order of  $1/m_A
$. When $m_{C}>>m_A$ the contribution of the mass polarization term can be
neglected \cite{FilKez2018}.  A physical result does not depend on
the reference frame. Thus, the MPT is not an artifact of using the reference frame associated with third particle. In the 
reference frame presented in Eq. (\ref{Sh}) this is a kinematic effect related
to the presence of the third particle $A$ when the other $A$ particle interacts with
the particle $C$. The presence of the third particle gives the
redistribution of kinetic energy, and as a result $AC$ subsystem is off the
energy shell \cite{FilKez2018}. If one considers the $AAC$ using Jacobi
coordinates by employing the Faddeev equations, the latter fact is hidden in
each Faddeev component that corresponds to the interaction of any two
particles in the presence of the third. 

In Table  \ref{tab-2R1} the contribution of the MPT is denoted as $\delta$. In the $AAC$ system, the  mass polarization effect is the following:
\begin{equation}
2E_2 -E_{3}(V_{AA}=0)-\delta=0.
\label{p1}
\end{equation}
The $\delta$ mainly depends on the mass ratio $m_A/m_C$ and $\Omega N$ interaction.  Analysis of the MPT in Table \ref{tab-2R1} shows that $\delta$ weakly depends on the set of the HAL QCD potential fitting parameters, less than 5\%. Consideration of the physical and unphysical $\Omega$ baryon and nucleon masses for the HAL QCD interaction also changes the MPT contribution at about 7\%. In contrast, in the case of the meson exchange potential Eq. (\ref{Hyodo}) the MPT contribution is about 30\% smaller than for the potential (\ref{HALInteraction}).

\subsection{From $AAC$ to $ABC$ model: Effects of the
Coulomb force}

A consideration of the Coulomb attraction makes three particles undistinguishable and requires a description of the $\Omega d$ in the framework of $ABC$ model. Within the theoretical formalism presented in the previous section we
calculate the ground state energy $E_{3}$ for the $\Omega d$, the two-body energies $E_{2}^{\Omega n}$ and $E_{2}^{\Omega p}$ of the
bound pairs and the three-particle interaction energy $E_{3}(V_{pn}=0)$ when the interaction between nucleons is omitted but they interact with $\Omega$ baryon \textit {i. e.} in the system presents only $\Omega N$ interactions.  The results of calculations of these energies for the $AAC$ and $ABC$ models are listed in Table \ref{tab-2R}.  Interestingly enough, the binding energy of $\Omega np$ is greater than that of the $^{3}$H. This is mainly because: i. both $\Omega N$ \cite{Iritani2019,LagrangianMethod} interactions are strongly attractive; ii. there is no Pauli exclusion principle limitation between $\Omega^{-}$ and nucleons; iii. a more massive $\Omega^{-}$ reduces kinetic energy, favoring a more bound state.

\begin{figure}[b]
\begin{center}
\includegraphics[width=14pc]{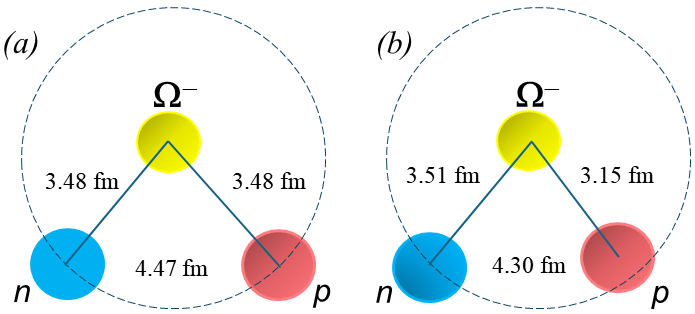}
\end{center}
\caption{The schematic representation of the $ABC$ system ($\Omega  pn$) when the nuclear $NN$ interaction is ignored: ($a$) The Coulomb interaction in $\Omega p$ pair is absent.
($b$) The pairs $(\Omega n)$ and $(\Omega p)$ are different due to the Coulomb interaction and bound with different two-body energies
$E_{2}^{\Omega^{-} n}$ and $E_{2}^{\Omega^{-} p}$. The numbers indicate the root mean square distances between the particles.
} \label{fig:40a}
\end{figure}

The analysis of the results in Table \ref{tab-2R} shows that the Coulomb attraction increases the $E_{3}$  and $E_{3}(V_{pn}=0)$ by about 0.9 MeV the HAL QCD potential. This increase does not depend either on $NN$ or the parameter sets for $\Omega N$ (\ref{HALInteraction}) interactions. Moreover, the Coulomb attraction leads to a more compact configuration of the $\Omega np$ tribaryon: $rms$ distances between particles decrease as illustrated in Fig. \ref{fig:40a}. In the case of meson exchange potential (\ref{Hyodo}), the contribution of the Coulomb interaction is $\sim$ 1.3 MeV. Thus, the description of the $\Omega np$ system using the meson exchange potential  (\ref{Hyodo}) leads to a more sizeable contribution of the Coulomb interaction to the $\Omega np$ binding energy.

Consideration of the Coulomb force leads to the $ABC$ model and changes Eq. (\ref{p1}) to the following one:
\begin{equation}
E_2(AC)+E_2(BC)-E_{3}(V_{AB}=0)-\delta=0.
\label{p2}
\end{equation}
The Coulomb force shifts of the two-body energy as $\Delta_{C}=E_2(AC)-E_2(BC)$ and gives the value of 0.9 MeV for the $V_{L1\Omega N}$ and 1.3 MeV for the $V_{Yu\Omega N}$ potentials, respectively. 

First, consider the influence of the MPT in the case of HAL QCD potential. The value of the mass-polarization is relatively small and can be assumed to be a constant
due to the predomination of the mass of the $\Omega$ particle. We found that the wave function of the $ABC$ system demonstrated a weak dependence on the Coulomb force. This is indicated by the localization of the particles in the system, which does not practically change after switching on the Coulomb potential (see Table \ref{t33}).
Due to the mass-polarization $\delta$ being a part of the matrix element of the kinetic energy operator, we can conclude that the Coulomb force does not change the $\delta$. Taking into account Eq. (\ref{p2}), one can see that the energy $E_{3}(V_{AB}=0)$ has to be changed following the two-body energy change.
From Eq. (\ref{p2}), it follows that the energy $E_{3}(V_{AB}=0)$ changed by the same amount as the two-body energy $E_2(BC)$ changed due to the Coulomb interaction.
The latter means the energy of the three-body  $\Omega d$ system is changed by the same value of 0.9 MeV listed in the Table. \ref{tab-2R}. 

\begin{table}[t]
\caption{ The energy characteristics of the $\Omega d$ %
 in $ABC$ model  with the Coulomb interaction and when the Coulomb interaction is neglected. Calculations performed with different $\Omega N$  and $NN$ potentials.
$E_{3}$ is the ground state energy of the $\Omega d$, $E_{2}({\Omega N})$ is the energy of bound $\Omega^{-} n$ and $\Omega^{-} p$ pairs, respectively, and $E_{3}(V_{NN}=0)$ is the
three-body energy, when the $NN$ interaction between nucleons is omitted. The energies are given in MeV.
The results of Refs. \cite{GV0,GV,Zhang2022,ESE2023} are shown for the comparison.  }
\label{tab-2R}\centering
\begin{tabular}{@{}lccccccccc}
\hline
\noalign{\smallskip} Potentials &
$E_{3}$&  $E_{3}$ \cite{GV0}&  $E_{3}$ \cite{GV}&$E_{3}$ \cite{Zhang2022}& $E_{3}$\cite{ESE2023} & $E_{3}(V_{NN}=0)$ & $E_{2}^{\Omega^{-} n}$ &$E_{2}^{\Omega^{-} p}$ &$\delta$\\ \hline 
\noalign{\smallskip}
 $V_{L1\Omega N}$, ATS3 (Coulomb)  & -20.8 &&&&& -3.86 &-1.2909& -2.1595& 0.41 \\
 $V_{L1\Omega N}$, ATS3 (no Coulomb)  & -19.9 & &&& &-2.99  & -1.2909& -1.2909 & 0.41\\
 $V_{L1\Omega N}$, MT (Coulomb)  & -20.5& &-20.9&& -20.935&-3.86  &  -1.2909&-2.1595 & 0.41\\
 $V_{L2\Omega N}$, MT (Coulomb)  & -20.9 && -21.3& -22.0& & -4.08 & -1.3759 & -2.2645&0.44\\
$V_{Yu\Omega N}$, MT (Coulomb) & -18.06 &-17.35 &&&& -2.16 & -0.3385  &-1.1787&0.64\\
$V_{Yu\Omega N}$, MT (no Coulomb) & -16.77 &-16.34 &&&& -0.97 & -0.3385  &-0.3385&0.30\\
\noalign{\smallskip}
\hline
\end{tabular}%
\end{table}
Here, it has to be mentioned that the results of the calculations presented for the $AAC$ model in Table \ref{tab-2R1} and those in Table \ref{tab-2R} for the $ABC$ model were obtained using different numerical procedures. The first method can be described as a direct numerical approach based on the finite-difference method. The numerical accuracy of this approach depends on the precision of the finite-difference approximation on a two-dimensional coordinate mesh. An example of using this method can be found in Ref.  \cite{FSV17}.
The second approach involves the reduction of the Faddeev equations to a set of one-dimensional equations by expanding the wave function on the bases of eigenfunctions of two-body Hamiltonians describing the subsystems of the three-body system  
\cite{Ya}. This expansion simplifies the computational problem but introduces an additional source of numerical errors.  
A brief description of the cluster reduction method can be found in Ref. \cite{KezerasPRD2020}. Taking this into account, we can highlight that the results in Table \ref{tab-2R1} were obtained with a better accuracy.  
\begin{table}[!ht]
\caption{
The root mean squarde distances in fm in the  $\Omega d$ system along with the ground state energy $E_{3}$ and the Coulomb energy shift $\Delta_{C}$ in MeV. The $V_{L1\Omega N}$ and $V_{L2\Omega N}$ potentials are used for the $\Omega N$ interaction and the MT \cite{Malfliet1969} potential for the nucleon-nucleon interaction.
}
\label{t34}
\begin{tabular}{lcccccc} \hline\noalign{\smallskip}
Model &  $d_{\Omega^- -p}$ & $d_{\Omega^- - n}$ & $d_{n-p} $ & $d_{\Omega^- - (np)} $& $E_3$, MeV & $\Delta_{C}$, MeV\\
\hline\noalign{\smallskip}
$ABC$(no Coulomb)  $V_{L1\Omega N}$  & 1.78 & 1.78 & 2.03& 1.47& -19.6& -- \\
   $ABC$(no Coulomb) $V_{NN}=0$  $V_{L1\Omega N}$  & 3.48 & 3.48 & 4.47& 2.66& -2.99& -- \\
      $ABC$(Coulomb) $V_{NN}=0$  $V_{L1\Omega N}$  & 3.51 & 3.15 & 4.30& 2.55& -3.86& -0.9 \\
$ABC$(no Coulomb)  $V_{L2\Omega N}$  & 1.77 & 1.77 & 2.02& 1.46& -20.0& -- \\
$AAC$(Coulomb) \cite{ESE2023} & 1.768 & 1.768 & 2.001& 1.458& -20.935&n/a\\
$ABC$(Coulomb) $V_{L1\Omega N}$  & 1.77 & 1.79 &2.03& 1.47& -20.5& -0.9\\
$ABC$(Coulomb) $V_{L2\Omega N}$  & 1.76 & 1.78 &2.02& 1.45& -20.9& -0.9\\
\noalign{\smallskip}\hline
\end{tabular}
\end{table}

We can see how works the algebraical relations of  Eq. (\ref{p1}). Let us use Eq. (\ref{p1}) and data from Table \ref{tab-2R} and  calculate mass polarization contribution in three-body system for two $\Omega N$ potentials:
$\delta_1 = -2\times 1.2909+2.997 = 0.4152$ MeV for $V_{L1\Omega N}$ and $\delta_2 = -2\times 1.3759+3.189 = 0.4372$ MeV for $V_{L2\Omega N}$ potentials. The difference of the mass polarization is equal approximately to $-0.02$ MeV. 
On the other hand,  we can estimate the two-body energies difference related to the $V_{L1\Omega N}$ and $V_{L2\Omega N}$ potentials as $E_2$(L1)-$E_2$(L2) $= (1.2909-1.3759) = -0.085$ MeV.
We have 0.19 MeV (see Table. \ref{tab-2R1}) for evaluation of the effect of the potential variation on the three-body energy $E_{3}(V_{NN=0})$: -2.997+3.189 MeV.
This value is compensated by the sum of the $E_2$ energy difference multiplied by two  ($-0.17$ MeV) and the difference of the mass polarizations ($-0.02$ MeV). Thus, in the case when the Coulomb interaction is included in the $BC$ pair, the mass polarization did not practically change and the three-body energy is approximately changed by the amount of the two-body Coulomb energy in the $BC$ pair.

To continue this study, we aim to demonstrate the impact of the Coulomb potential on the $ABC$ system and compare our results with those for the $AAC$ model.  
We assume that the Coulomb potential introduces a perturbation in the $AAC$ system.  
The spatial configurations of the $ABC$ and $AAC$ systems differ due to the symmetry violation in the $\Omega^- n$ and $\Omega^- p$ pairs. The results of our evaluation for the spatial configurations of the $\Omega^- np$ system, based on different potential models, are presented in Table \ref{t34}.  
We provide the root mean square distances between particles, calculated using Eq. (\ref{e:dist}), for the $V_{L1\Omega N}$ and $V_{L2\Omega N}$ potentials in the two bottom rows. The results are identical due to the similarity in binding energies.  
More interesting results are obtained for the model where the nucleon-nucleon potential is switched off (the second and third rows in Table \ref{t34}). The Coulomb potential breaks the isosceles triangle symmetry of the $AAC$ model, as illustrated in Fig. \ref{fig:40a}.  
Including the nucleon-nucleon interaction increases the compactness of the system, thereby masking the symmetry violation caused by the Coulomb potential. These effects can be attributed to the dominance of the strong $\Omega N$ interaction. 
Thus, our results for the $ABC$ model are in good agreement with those for the $AAC$ model published in Ref. \cite{ESE2023}.  

\begin{figure}[t]
\begin{center}
\includegraphics[width=15pc]{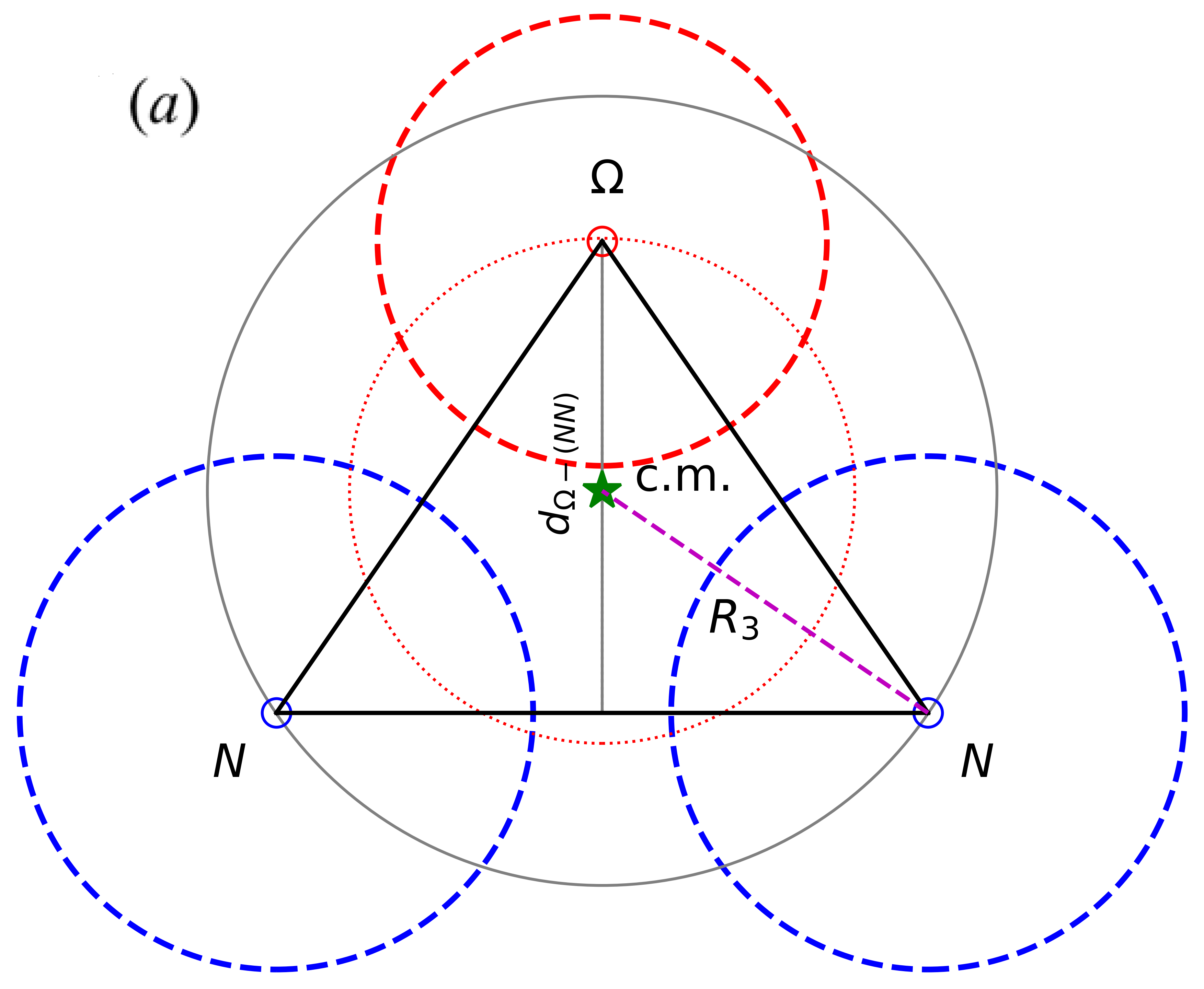}
\ \ \ \ \ \ \
\includegraphics[width=15pc]{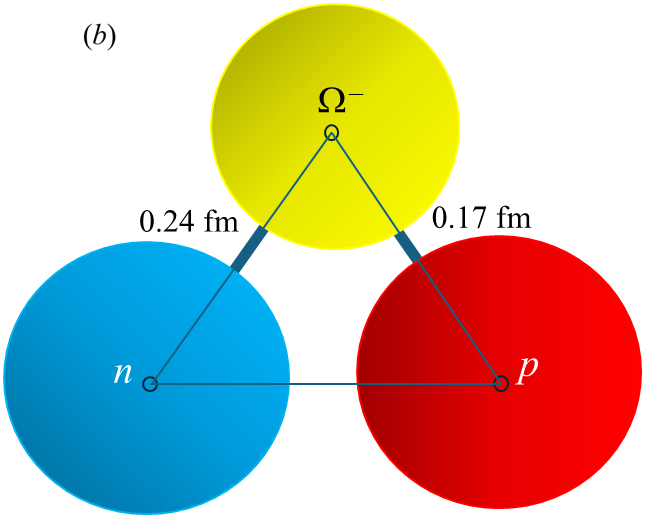}

\end{center}
\caption{The schematic representation of the $\Omega^-np$ system. In these calculations, we use  $V_{L1\Omega N}$ and MT potentials. ($a$) The Coulomb potential is not considered, reducing the system to $\Omega NN$. The star marks the center-of-mass of the system (c.m.). The particle sizes are set to 0.8 fm for nucleons and 0.7 fm for the $\Omega$ baryon, indicated by dashed circles. The vertical line represents the $rms$ distance between the $\Omega$ baryon and the center-of-mass of the nucleon pair.   ($b$) The Coulomb potential is included, breaking the isosceles triangle symmetry in the $\Omega^-np$ system. This asymmetry can be observed through differences in the space gap (bulk solid line) between "sufaces" of $\Omega^-$ baryon and the proton and neutron. Numbers display these differences. Here, the effective radius of the proton is 0.83~fm}
     \label{fig:40b}
\end{figure}

The spacial configuration of particles in the $\Omega NN$ system are shown in Fig. \ref{fig:40b}($a$) for the case of the $V_{L1\Omega N}$ potential. 
To evaluate this configuration, we take into account the Coulomb interaction.
It can be seen that the particles are located close to one another, especially when the distance is given as the separation between their effective "surfaces." In this context, we assume effective particle radii of about 0.7–0.9 fm. The reason for this compactness and large binding energy is the deeply attractive core of the $\Omega N$ potential near the origin. The short to medium-range behavior of the potential at short distances plays a crucial role, resulting in a substantial binding energy due to the significant depth of the attractive core.
A slight geometric asymmetry arises due to the Coulomb interaction between the $\Omega^-$ and the proton as it is shown in Fig.  \ref{fig:40b}($b$).

Within the $AAC$ model, assuming the particles are spherically symmetric with radii $R_\Omega$,  $R_N$, $R_N, $ and are located at positions \( \mathbf{r}_1, \mathbf{r}_2, \mathbf{r}_3 \) with respect to an origin, and the center-of-mass of the system is at \( \mathbf{R}_{\text{cm}} \), the mean square radius of the system can be given by:
\begin{equation}
\langle R^2 \rangle = \frac{1}{M} \sum_{i=1}^{3} m_i |\mathbf{r}_i - \mathbf{R}_{\text{cm}}|^2 + \frac{1}{M} \sum_{i=1}^{3} m_i \langle r^2 \rangle_{i,\text{internal}},
\label{r1}
\end{equation}
where
\( m_i \) is the mass of the \( i \)-th particle.
   \( M = \sum_{i=1}^{3} m_i \) is the total mass of the system.
 \( |\mathbf{r}_i - \mathbf{R}_{\text{cm}}|^2 \) is the squared distance of the \( i \)-th particle from the center of mass.
 \( \langle r^2 \rangle_{i,\text{internal}} \) is the $rms$ of the \( i \)-th particle with respect to its own center.  
For the model including two nucleons and the $\Omega$ baryon having masses \( m_N \) and  \( m_{\Omega} \), the formula Eq. (\ref{r1}) simplifies to:
\begin{equation}
\langle R^2 \rangle = \frac{1}{2m_N + m_{\Omega}} ( m_N |\mathbf{R}_3 |^2 + m_N |\mathbf{R}_2|^2 + m_{\Omega} |\mathbf{R}_1 |^2  +
2 m_N R_N^2  + m_{\Omega} R_{\Omega}^2 ),
\label{r2}
\end{equation}
where 
$\mathbf{R}_i=\mathbf{r}_i - \mathbf{R}_{\text{cm}}$, $i=1,2,3$.
The root mean square radius of the $\Omega NN$ system is then: $R_{\text{rms}} = \sqrt{\langle R^2 \rangle}$.
To calculate $R_{\text{rms}}$, we used the  values $R_N$=0.8 fm and  $R_{\Omega}$=0.7 fm, that is shown in Fig.  \ref{fig:40b}($a$).
This input leads to the $rms$ of the $\Omega NN$ about 1.29 fm and sligthly disagrees to the value for the matter $rms$ radius of 1.097~fm reporeted in Ref. \cite{ESE2023} due to differenet definitions for the values.

\section{Summary and Concluding Remarks}
\label{conclusion}

In this study, we have investigated the $\Omega^{-} d$ exotic nucleus using the HAL QCD $\Omega N$ potential and the $\Omega N$  potential based on a baryon-baryon interaction model with meson exchanges. Both interactions indicate the existence of $\Omega N(5/2^{+})$ bound state. 
Based on the Faddeev equations in configuration space, we examine the \(\Omega N\) potentials using two models. The first treats the \(\Omega^- np\) system as a three-body system with two identical particles (the $AAC$ model). The second model incorporates the attractive Coulomb force between the \(\Omega^-\) and proton, treating the system as composed of three non-identical particles (the $ABC$ model). Numerical analysis of the $ABC$ ($ACC$) model was performed using the cluster reduction method~\cite{Ya,KezerasPRD2020} (direct finite-difference method, \cite{FSV17}).

Both $\Omega N$ potentials lead to the bound state of the $\Omega d$  system ($^3_{\Omega}$H nucleus) with binding energy 19.58 MeV and 16.77 MeV with the HAL QCD \cite{Iritani2019} and meson exchange \cite{LagrangianMethod} interactions, respectively, when the Coulomb interaction is not considered. The Coulomb interaction increases the binding energy and changes the results to 20.5 MeV and 18.06 MeV, respectively, and its contribution is $\sim$ 30\% larger in the case of meson exchange potential \cite{LagrangianMethod}.   

Our goal was to examine the \(\Omega N\) potentials within the three-body \(\Omega NN\) system. Although both \(\Omega N\) potentials produce only a weakly bound \(\Omega N\) pair with a binding energy of less than 1.3~MeV, the three-body \(\Omega^{-} d\) system exhibits strongly bound states with binding energies nearly ten times larger. An analogy to the nucleon-nucleon interaction is not appropriate, as the ratio of two- to three-body binding energies differs significantly$-$by a factor of approximately 3.8$-$when comparing the deuteron (\(\sim 2.22\)~MeV) and the triton (\(\sim 8.482\)~MeV). Clearly, a key difference between the \(NN\) and \(\Omega N\) interactions is the presence of a repulsive core in the nucleon-nucleon force, which limits the binding energy in the three-nucleon system.
In contrast, both \(\Omega N\) potentials exhibit a strongly attractive core at short distances. This strong attraction leads to a rapid increase in the three-body binding energy when the attractive nucleon-nucleon interaction is switched on, resulting in a compact three-body spacial configuration as the \(\Omega N\) distances become small.

The low-energy characteristics of $\Omega N$ and $\Omega^{-} d$ systems weakly depend on the fitting parameter sets for the HAL QCD potential. 
The comparison of the results obtained with the $\Omega N$ potentials of  Eqs. (\ref{HALInteraction}) and (\ref{Hyodo}) show significant differences for low-energy characteristics. Thus, the low-energy parameters of these systems are very sensitive to the form of the $\Omega N$ potential.

We evaluate the effect of the Coulomb force between the proton and  $\Omega^-$ baryon in the $ABC$ model. We have shown that the strong $\Omega N$ potential defines the spatial structure of the system and renders the influence of the Coulomb force small. 
Our predictions for the binding energy differ slightly from previously published results. However, they qualitatively agree with these earlier findings and indicate the existence of bound or quasi-bound exotic systems. In this study, we assess the impact of the Coulomb interaction on the system and compare our results with those from $AAC$ calculations obtained using various approaches, such as integral Faddeev equations and the variational method. The $ABC$ model leads to the low-energy characteristics for the $\Omega NN$ system that differ from previous calculations.
The Coulomb potential has a marginal perturbative effect on the $AAC$ system, shifting the three-body binding energy by the Coulomb energy value obtained in the two-body $BC$ subsystem but slightly deviating the spatial configuration from isosceles triangle symmetry. These effects are primarily driven by the  strong attractive $\Omega N$ interaction.

The $\Omega N$ interaction is predicted to be strongly attractive in the $^{5}S_2$ channel. Let us speculate about the binding energy per baryon in $\Omega$-baryonic nuclei. 
In ordinary atomic nuclei, the binding energy per nucleon doesn't keep increasing endlessly by combining additional nucleons due to the strong short-range $NN$ repulsion, the Pauli principle preventing nucleons from crowding into the lowest energy state, and the Coulomb repulsion between protons that grow with $Z$. These lead to the peak limitation of the binding energy per nucleon $\sim 8.9$ MeV for iron/nickel. Combining $\Omega^{-}$ baryons to original nuclei is completely different from adding nucleons because $\Omega N$ interaction is attractive and $\Omega^{-}$ doesn’t obey Pauli exclusion with nucleons. The presence of $\Omega^{-}$  deepens the potential well felt by the nucleons, and each nucleon can be more tightly bound than in ordinary nuclei.  The binding energy per baryon may exceed 8 MeV, especially in a few nucleon systems with $\Omega^{-}$. When we added a single $\Omega^{-}$ to a deuteron, the strong $\Omega N$ interaction  significantly enhances the binding energy per baryon above the typical 2.6 and 2.8 MeV/nucleon seen in $^{3}$H and $^{3}$He nuclei, respectively. We can hypothesize that as more nucleons are added to $\Omega^{-}np$, the system should reach a saturation level - but at a higher binding energy per baryon, maybe around 12-15 MeV. This is because the $\Omega^{-}$ contributes additional binding due to the increase of ($\Omega N$) pairs without being subject to the Pauli exclusion principle with the nucleons. If the system contains more than one $\Omega^{-}$, the saturation level might be pushed even further upward because the additional attractive $\Omega \Omega$ interaction \cite{Gongyo2018} further deepens the potential well. 

In summary, the binding energy of the $\Omega^{-} d$ depends on the details of the $\Omega N$ potential. Our calculations suggest that the system is quite tightly bound, but due to the high mass of the $\Omega$ baryon, the system could be unstable, decaying into other particles, especially if the interaction is not strong enough to counteract the decays of the $\Omega$ baryon.
Although a $\Omega d$ bound state is theoretically possible, its stability would be an issue due to the short lifetime of the $\Omega$ baryon. 
While stable nuclei consisting solely of $\Omega$ baryon and nucleons may be unlikely due to the short lifetime of the $\Omega$ baryon typical on the order of $\sim$ 0.1 ns 
and the complex nature of their interactions, this idea fits within the broader context of strange matter. The presence of $\Omega$ baryons would introduce additional strangeness to the system. Typically, systems with more strange quarks are expected to have different properties compared to normal matter. There are predictions that strange quark matter, which could consist of strange quarks, up quarks, and down quarks, might be stable at very high densities, such as in the interior of neutron stars \cite{Witten1984,Jaffe1984,Zhang2024}. However, a nucleus consisting of $\Omega$ baryons and nucleons would need to be stable at much lower densities, potentially as an exotic form of matter.   Further theoretical models, experimental research, and simulations, such as those using lattice QCD, are needed to better understand the feasibility, stability, and structure of such exotic nuclear systems. 
Nonetheless, such systems provide an exciting frontier for both theoretical nuclear physics and astrophysics, especially in understanding the role of strange quarks in dense nuclear matter.  Further theoretical models, experimental research, and simulations, such as those using lattice QCD, are needed to better understand the feasibility, stability, and structure of such exotic nuclear systems. Our study indicates the existence of bound or quasi-bound exotic
systems, providing a guideline for future experimental searches.

\section*{Acknowledgments}

This work is supported by the National Science Foundation grant HRD-1345219
and DMR-1523617 awards Department of Energy$/$National Nuclear Security
Administration under Award Number NA0003979 DOD-ARO grant \#W911NF-13-0165.

\end{document}